\documentclass[aps,prl,twocolumn,groupedaddress,showpacs]{revtex4}

  \usepackage{graphicx}

\begin{document}


\title{Controlling Fast Chaos in Delay Dynamical Systems}




\author{Jonathan N. Blakely}
\altaffiliation{present address: US Army Research Development and Engineering Command, AMSRD-AMR-WS-ST, Redstone Arsenal, Alabama 35898}

\author{Lucas Illing}
\email[]{illing@phy.duke.edu}

\author{Daniel J. Gauthier}
\affiliation{Department of Physics and Center for Nonlinear and Complex
Systems, Duke University, Durham, North Carolina, 27708, USA}

\date{\today}
\begin{abstract}
We introduce a novel approach for controlling fast chaos in time-delay dynamical systems and use it to control a chaotic photonic device with a characteristic time scale of $\sim $12 ns. Our approach is a prescription for how to implement existing chaos control algorithms in a way that exploits the system's inherent time-delay and allows control even in the presence of substantial control-loop latency (the finite time it takes signals to propagate through the components in the controller). This research paves the way for applications exploiting fast control of chaos, such as chaos-based communication schemes and stabilizing the behavior of ultrafast lasers.
\end{abstract}


%

\maketitle
 
Because of its unpredictable nature, chaos is often viewed as an undesirable characteristic of practical devices. However, recent studies have shown that chaos can be used for a variety of applications such as information transmission with high power efficiency \cite{Ott_Chaos}, generating truly random numbers \cite{gleeson,kocarev}, and novel spread spectrum \cite{kennedy}, ultrawide-bandwidth \cite{rulkov_cppm,rulkov_uwb}, and optical \cite{Roy_Science} communication schemes. For many of these applications, it is desirable to operate the devices in the fast regime where the typical time scale of the chaotic fluctuations is on the order of 1 ns \cite{Roy_Science,illing_QE}.  Many applications also require the ability to control the chaotic trajectory to specific regions in phase space \cite{Hayes_PRL}. 

On the fast scale, the time it takes for signals to propagate through the device components is comparable to the time scale of the fluctuations and hence many fast systems are most accurately described by time-delay differential equations. 
Examples of fast broadband chaotic oscillators that are modeled as time-delay systems include electronic \cite{Mykolaitis},  opto-electronic \cite{illing_QE,Goedgebuer_PRL}, and microwave oscillators \cite{Ott_Chaos}, as well as lasers with delayed optical feedback \cite{Roy_Science}, and nonlinear optical resonators \cite{Ikeda_PRL}.
An advantageous feature of these time-delay devices is that the complexity of the dynamics can be tuned by adjusting the delay \cite{farmer}.

For applications requiring controlled trajectories, it is possible to use recently developed chaos-control methods.
The key idea underlying these techniques is to stabilize a desired dynamical behavior by applying feedback through minute perturbations to an accessible parameter when the system is in a neighborhood of the desired trajectory in state-space \cite{Gauthier_AJP,OGY}.
In particular, many of the control protocols attempt to stabilize one of the unstable periodic orbits (UPOs) that are embedded in the chaotic attractor.
Although the research has been very successful for slow systems (characteristic time scale $>$ 1 $\mu $s) \cite{Roy_PRL,Hunt_PRL,Petrov_Nature,Garfinkel_Science}, applying feedback control to fast chaotic systems is challenging because the controller requires a finite time to sense the current state of the system, determine the appropriate perturbation, and apply it to the system. 
This finite time interval, often called the control-loop latency $\tau _{\ell }$, can be problematic if the state of the system is no longer correlated with its measured state at the time when the perturbation is applied.
Typically, chaos control fails when the latency is on the order of the period of the UPO to be stabilized \cite{Just_PRE,Hoevel_PRE,Sukow_Chaos}.

The control of very fast chaotic systems is an outstanding problem
because of two challenges that arise: control-loop latencies are unavoidable, and complex high-dimensional behavior of systems is common due to inherent time-delays. The primary purpose of this Letter is to describe a novel approach for controlling time-delay systems even in the presence of substantial control-loop latency. This is a general approach that can be applied to any of the fast systems described above. As a specific example, we demonstrate this general approach by using time-delay autosynchronization control \cite{Pyragas} to stabilize fast periodic oscillations ($T_{PO} \sim $12 ns) in a photonic device, the fastest controlled chaotic system to date \cite{Corron_IJBC}. In principle, much faster oscillations can be controlled using, for example, high speed electronic or all-optical control components, paving the way to using controlled chaotic devices in high-bandwidth applications \cite{Hayes_PRL,comment1}.

%
%
%
\begin{figure}
\includegraphics[height=\columnwidth,angle=90]{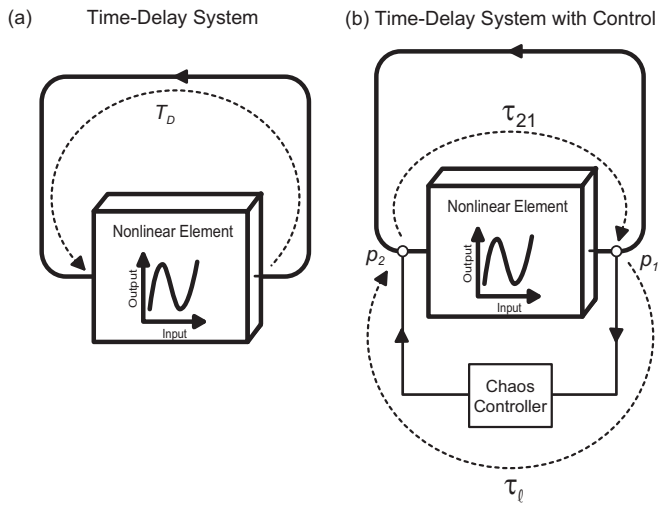}
\caption{ (a) Schematic of a typical topology of a time-delay system consisting of a nonlinear element and a long loop connecting the output to the input of the element, introducing a total time-delay $T_D$. (b) Schematic of feedback control that measures the state at point $p_{1}$ and perturbs the system at $p_{2}$. The propagation time through the controller (control loop latency) is denoted by $\tau_{\ell}$ and $\tau_{21}$ is the propagation time from $p_{2}$ to $p_{1}$. Note that the signal takes a time $T_D-\tau_{21}$ to get from  $p_{1}$ to $p_{2}$. }
\label{fig:schematic_setup}
\end{figure}

We have discovered that the effects of control-loop latency can be mitigated when controlling chaotic systems involving a nonlinear element and an inherent time-delay $T_{D}$, as shown schematically in 
Fig.~\ref{fig:schematic_setup}a. 
Chaos can be controlled in time-delay systems by taking advantage of the fact that it is often possible to measure the state of the system at one point in the time-delay loop ($p_{1}$) and to apply perturbations at a different point ($p_{2}$), as shown in Fig.~\ref{fig:schematic_setup}b. 
Such distributed feedback is effective because the state of the system at $p_{2}$ is just equal to its state $p_{1}$ delayed by the propagation time $T_D-\tau_{21}$ through the loop between the points.
The arrival of the control perturbations at $p_{2}$ is timed correctly if 
\begin{equation}
\label{timing}
\tau_{\ell}+\tau_{21}=T_{D}.
\end{equation}
  Hence, it is possible to compensate for a
reasonable amount of control loop latency $\tau_{\ell}$ by appropriate choice of $p_{1}$
and $p_{2}$. The advantage of this approach is that the propagation time
through the controller does not have to be faster than the controlled
dynamics. In contrast, the conventional approach for controlling chaos is
to perform the measurement and apply the perturbations
instantaneously ($\tau_{\ell} \rightarrow 0$), which requires controller components that are much faster than the components of the chaotic device to approximate instantaneous feedback.
Note that we have not specified a method of computing the control perturbations. In principle, 
any of the existing methods \cite{Gauthier_AJP} may be used as long as they can be implemented with 
latency satisfying Eq. \ref{timing}.

%
%
%
\begin{figure}
\includegraphics*[height=\columnwidth,angle=90]{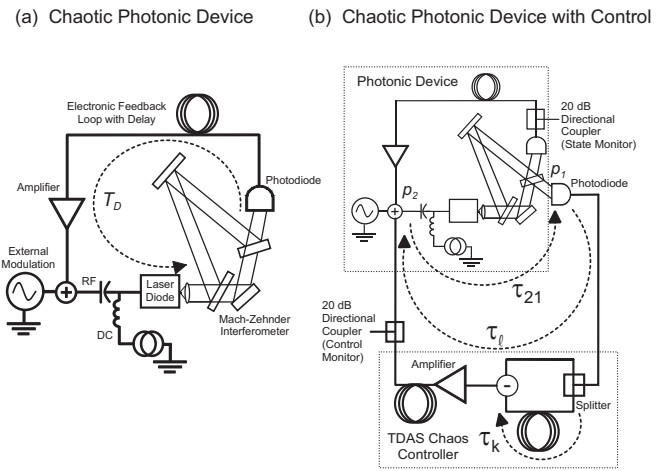}
\caption{(a) Experimental setup of a chaotic time-delay device of the type shown in Fig.~\ref{fig:schematic_setup}a. The nonlinear element is a Mach-Zehnder interferometer and the delay loop consists of a photodiode, coaxial cable, an amplifier and a semiconductor laser. The electronic signal is applied to the laser through a bias-T (capacitor and inductor) which converts the RF-signal to a current and adds it to the DC-injection current. See \cite{Blakely_QE} for details. 
(b) Experimental system with controller.
The measurement point $p_{1}$ is the second beam splitter of the interferometer. Perturbations are applied at $p_{2}$, an RF-power combiner.  
The time $\tau_{21}$ for a signal to propagate from point $p_{2}$ to $p_{1}$ is $\sim$ 3 ns.
The controller contains two delay lines, the first sets $\tau_{k}$, the period of the orbit to be controlled, the second is used to adjust the latency $\tau_{\ell}$ to properly time the arrival of perturbations at $p_{2}$. The state of the system is monitored through a directional coupler positioned directly after the photodiode in the delay loop of the photonic device. The control signal is measured through a directional coupler at the output of the controller.
}
\label{fig:experimental_setup}
\end{figure}

To demonstrate the feasibility of controlling fast chaos using this general
concept, we apply it to a chaotic photonic device shown schematically in
Fig.~\ref{fig:experimental_setup}a. The device consists of commercially-available components
including a semiconductor laser, a Mach-Zehnder interferometer,
and electronic time-delay feedback, and can display nanosecond-scale chaotic
fluctuations \cite{Blakely_QE}. The semiconductor laser acts as simple current-controlled
source that converts current oscillations into oscillations of the optical
frequency and, to a lesser extent, amplitude. Light generated by the laser traverses an
unequal-path Mach-Zehnder interferometer whose output is a nonlinear
function of the optical frequency.  The light exiting the interferometer is
converted to a voltage using a fast silicon photodiode and a resistor. This voltage
propagates through a delay line (a short piece of coaxial cable), is
amplified and bandpass filtered, and is then used to modulate the laser
injection current by combining it with the dc injection current via a bias-T. 
The system is subject to an external driving force provided by adding an 
RF voltage to the feedback signal.
(The driven system has more prominent bifurcations than the undriven device.)
The length of the coaxial cable can be adjusted to obtain values of the time-delay $T_D$ in the range 11 - 20 ns. 
This photonic device displays a range of
periodic and chaotic behavior that is set by the amplifier gain and the
ratio of the time-delay to the characteristic response time of the system
(typically set to a large value).

\begin{figure}
\includegraphics*[width=\columnwidth]{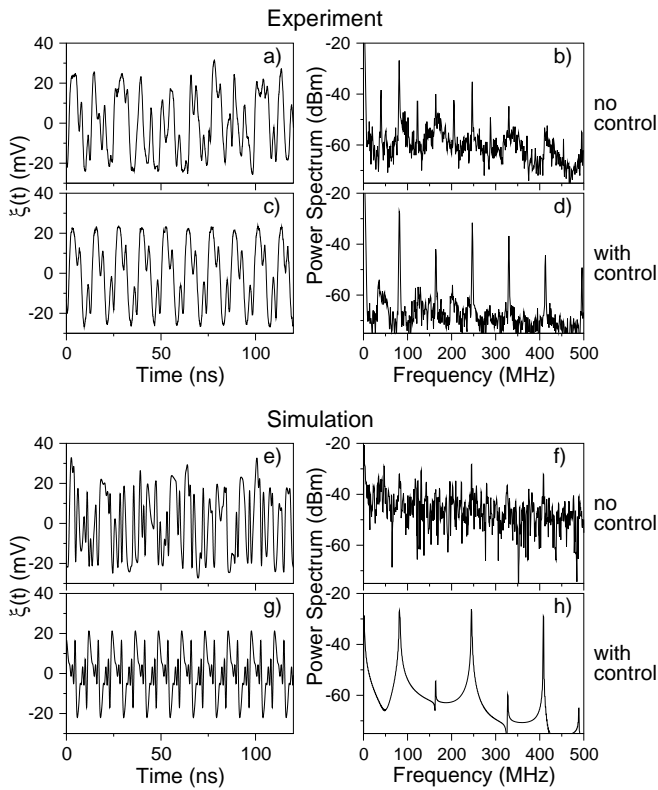}
\caption{ Experimental (a-d) and simulated (e-h) data showing control of fast chaos. The state of the system is monitored by measuring the voltage in the delay loop before the amplifier (see Fig. \ref{fig:experimental_setup}b). 
(a) The chaotic time series of the monitored voltage in the absence of control, (b) the corresponding broad power spectrum, (c) the periodic time series of the stabilized orbit with control on, and (d) the corresponding power spectrum.
 The effect of control in simulations is consistent with our experimental results, as shown by the simulated time series of the monitor voltage without (e) and with (g) control and the corresponding power spectra (f) and (h).
}
\label{fig:control}
\end{figure}

 Figure \ref{fig:control}a shows the chaotic temporal evolution of the voltage 
measured immediately
after the photodiode when the device operates in
the absence of control. The corresponding broadband power spectrum is shown in 
Fig.~\ref{fig:control}b. The behavior of the system is well described 
by a delay-differential equation, which we use to investigate numerically 
the observed dynamics \cite{Blakely_QE}. The
predicted chaotic oscillations and power spectrum are
shown in Fig.~\ref{fig:control}e and f, respectively. 

We apply our control method to the photonic device using the setup shown in Fig.~\ref{fig:experimental_setup}b by measuring the state of the system (denoted by $\xi (t)$) at point $p_{1}$ and injecting continuously a control signal  $\varepsilon (t+\tau _{\ell })$ at point $p_{2}$.
For a given device time-delay $T_D$, coaxial cable can be added or removed from the control loop to obtain a value of $\tau_{\ell}+\tau _{21}$ satisfying Eq.~\ref{timing}.
 To compute the control perturbations  $\varepsilon$, we use a scheme known as time-delay autosynchronization (TDAS), an algorithm that synchronizes the system to its state one orbital period in the past by setting
 $\varepsilon(t)=\gamma \lbrack \xi (t)-\xi (t-\tau _{k})]$ where
 $\tau _{k}$ is a control-loop delay that is set equal to the period $T_{PO}$ of the desired orbit, and $\gamma $ is the control gain \cite{Pyragas,Gauthier_PRE}.
When synchronization with the delayed state is successful, the trajectory of the controlled system is precisely on the UPO and the control signal is comparable to the noise level in the system. 
We emphasize that TDAS-control was chosen for ease of implementation in this proof-of-concept experiment but that our approach is consistent with other control methods applicable to delay systems (e.g.  \cite{ETDAS,Michiels}).

%
%
%

Controlling the fast photonic device is initiated by setting the various
control-loop time delays ($\tau_{\ell}+\tau _{21}$ and $\tau_k$) and applying
 the output of the TDAS controller to point $p_{2}$ with 
$\gamma$ set to a low value ($\gamma=$ 0.1 mV/mW). 
Upon increasing $\gamma$ to 10.3 mV/mW, 
we observe that 
$\varepsilon (t)$ decreases, which we further minimize by making fine adjustments
to $\tau _{k}$ and $\tau _{\ell }+\tau _{21}$.  Successful control is
indicated when $\varepsilon (t)$ drops to the noise level of the device. 
Figure \ref{fig:control}c shows the periodic temporal evolution  of the
controlled orbit with a period of $T_{PO} \sim $12 ns. 
The corresponding power spectrum, shown in Fig. \ref{fig:control}d, is dominated by a single fundamental frequency of 81 MHz and it's harmonics.
 The observation of successful stabilization of one of the UPOs embedded in the chaos of the uncontrolled photonic device is consistent with the theoretical prediction of a mathematical model describing the photonic device in the presence of control, as shown in Figs.~\ref{fig:control}g and h, where the simulated time series and power spectrum, respectively, indicate periodic oscillations.

The data shown in Fig. \ref{fig:control} is the primary result of our
experiment, demonstrating the feasibility of controlling chaos in
high-bandwidth systems even when the latency is comparable to the
characteristic time scales of the chaotic device (compare $T_{PO} \sim $12 ns and $\tau _{\ell }\sim $
8 ns).

%
%
%
\begin{figure}
\includegraphics*[width=\columnwidth]{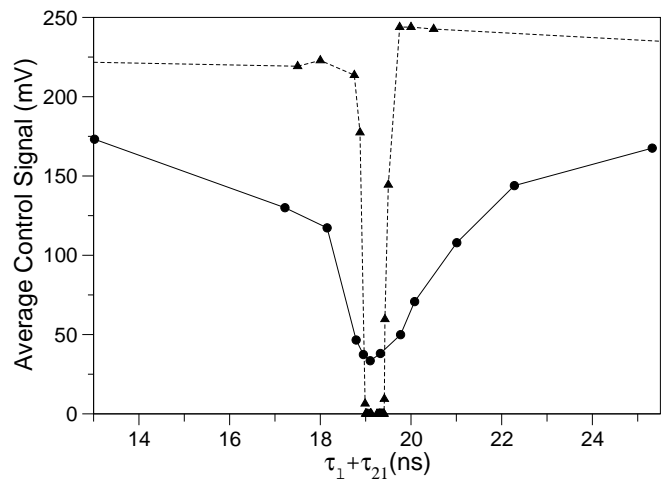} 
\caption{ Time-averaged control signal in the experiment (circles) and simulation (triangles). The minimum centered around $T_D=19.1$ ns is the region of successful control, where $\langle \varepsilon(t) \rangle$ is at the noise level ($\sim$ 30 mV, estimated by breaking the control loop and  measuring the control signal when the photonic device is in a periodic regime). 
 The width of the minimum ($\sim 0.5$ ns) indicates that control succeeds despite small errors in $\tau_{\ell}+\tau_{21}$. Noise in the experiment smooths out the sharp transition from controlled to uncontrolled behavior observed in simulation. }
\label{fig:robustness}
\end{figure}

To control this fast UPO, we used the smallest value of $\tau _{\ell }+\tau _{21}$
attainable with our current experimental apparatus.  Hence, it is not
possible to fully explore the effects on control when we change $\tau _{21}$. Therefore, we slowed down the chaotic photonic device by increasing 
$T_{D}$. 
 In this way, we can explore ($\tau _{\ell }+\tau _{21}$) over a range including values that are shorter than $T_{D}$. 
 Figure \ref{fig:robustness}
shows the size of the measured (circles) and predicted (line) control
perturbations as a function of $\tau _{\ell }+\tau _{21}$. \ It is seen that
control is possible over a reasonably large range of time delays ($\sim $0.5
ns) centered on $T_{D}$ so that it is not necessary to set precisely the
control-loop delay, a practical benefit of this scheme.

From the data shown in Fig.~\ref{fig:robustness}, we can infer what would happen if $p_{1}=p_{2}$ (the conventional method of implementing chaos control with nearly instantaneous feedback). In this case, $\tau _{21}=T_{D}$ and hence control would only be effective when $\tau_{\ell } < $0.5 ns, which is not possible using our implementation of TDAS. Note that the shortest reported control-loop latency of a chaos controller is 4.4 ns \cite{Corron_PRL}, much too large to control our device.

In principle, faster time-delay chaotic systems can be controlled using our
approach as long as the controller uses technology (e.g., integrated circuits,
all-optical) that is as fast as the system to be controlled so that $\tau
_{\ell }$ is comparable to $T_{D}$. Traditional chaos control schemes
require that $\tau _{\ell }$ be much shorter than $T_{D}$,
increasing substantially the cost and complexity of the controller.  With
regards to potential applications, we note that adjustments to our
controller allows for controlling different UPOs embedded in the chaotic
system, which could be used for symbolic-dynamic-based communication
schemes. Overall, our research points out the importance of using time-delay
dynamical systems combined with distributed control for applications
requiring fast controlled chaos.

Finally, our approach of control in the presence of control-loop latency is equally useful for non-chaotic fast and ultrafast time-delay devices, where the fast time scale makes the suppression of undesired instabilities challenging (e.g. the double pulsing instability in femtosecond fiber lasers \cite{Wise}.)

\begin{acknowledgments}
 The authors thank J. E. S. Socolar for helpful discussions. This work was supported by the US Army Research Office (grant \# DAAD19-99-1-0199).
\end{acknowledgments}


\begin{thebibliography}{99}


\bibitem{Ott_Chaos} V. Dronov, M. R. Hendrey, T. M. Antonsen, Jr., and E. Ott, Chaos {\bf 14}, 30 (2004).

\bibitem{gleeson} J. T. Gleeson, Appl. Phys. Lett. {\bf 81}, 1949 (2002).

\bibitem{kocarev} T. Stojanovski, J. Pihl, and L. Kocarev,  IEEE Trans. Circuits Syst. --I: Fundam. Theor. Appl. {\bf 48}, 382 (2001).

\bibitem{kennedy} M. P. Kennedy, G. Kolumb\'{a}n, G. Kis, and Z. J\'{a}k\'{o}, IEEE Trans. Circuits Syst.--I: Fundam. Theor. Appl. {\bf 47}, 1702 (2000).

\bibitem{rulkov_cppm} N. F. Rulkov, M. M. Sushchik, L. S. Tsimring, and A. R. Volkovskii, IEEE Trans. Circuits Syst.--I: Fundam. Theor. Appl. {\bf 48}, 1436 (2001).

\bibitem{rulkov_uwb} G. M. Maggio, N. F. Rulkov, and L. Reggiani, IEEE Trans. Circuits Syst.--I: Fundam. Theor. Appl. {\bf 48}, 1424 (2001).

\bibitem{Roy_Science} G. D. VanWiggeren and R. Roy, Science {\bf 279}, 1198 (1998). 

\bibitem{illing_QE} H. D. I. Abarbanel, M. B. Kennel, L. Illing, S. Tang, H. F. Chen, and J. M. Liu, IEEE J. Quantum. Electron. {\bf 37}, 1301 (2001). 
 
\bibitem{Hayes_PRL} S. Hayes, C. Grebogi, and E. Ott, Phys. Rev. Lett. {\bf 70}, 3031 (1993).

\bibitem{Mykolaitis} G. Mykolaitis \emph{et al.}, Chaos Solitons Fractals {\bf 17}, 343 (2003).

\bibitem{Goedgebuer_PRL} J. P. Goedgebuer, L. Larger, and H. Porte, Phys. Rev. Lett. {\bf 80}, 2249 (1998).     

\bibitem{Ikeda_PRL} K. Ikeda, K. Kondo, and O. Akimoto, Phys. Rev. Lett. {\bf 49}, 1467 (1982).

\bibitem{farmer} J. D. Farmer, Physica D, {\bf 4D}, 366 (1982)

\bibitem{Gauthier_AJP} D. J. Gauthier,  Am. J. Phys. {\bf 71}, 750 (2003).

\bibitem{OGY} E. Ott, C. Grebogi, and J. A. Yorke,  Phys. Rev. Lett. {\bf 64}, 1196 (1990).

\bibitem{Roy_PRL} R. Roy, T. W. Murphy, Jr., T. D. Maier, Z. Gills, and E. R. Hunt, Phys. Rev. Lett. {\bf 68}, 1259 (1992).

\bibitem{Hunt_PRL} E. R. Hunt, Phys. Rev. Lett. {\bf 67}, 1953 (1991).

\bibitem{Petrov_Nature} V. Petrov, V. Gaspar, J. Masere, and K. Showalter, Nature {\bf 361}, 240 (1993).

\bibitem{Garfinkel_Science} A. Garfinkel, M. L. Spano, W. L. Ditto, and J. N. Weiss, Science {\bf 257}, 1230 (1992). 

\bibitem{Just_PRE} W. Just, D. Reckwerth, E. Reibold, and H. Benner, Phys. Rev. E {\bf 59}, 2826 (1999).

\bibitem{Hoevel_PRE} P. H\"ovel, and J. E. S. Socolar, Phys. Rev. E Phys. Rev. E {\bf 68}, 036206 (2003). 

\bibitem{Sukow_Chaos} D. W. Sukow, M. E. Bleich, D. J. Gauthier, and J. E. S. Socolar, Chaos {\bf 7}, 560 (1997). 

\bibitem{Pyragas} K. Pyragas, Phys. Lett. A {\bf 170}, 412 (1992).

\bibitem{Corron_IJBC} N. J. Corron, B. A. Hopper, and S. D. Pethel, Int. J. Bif. Chaos {\bf 13}, 957 (2003).

\bibitem{comment1} One such application is chaos communication, where a powerful way to implement information transfer is to encode information in the symbolic dynamics by controlling the chaos \cite{Hayes_PRL}. A specific example of such a device in the microwave regime is proposed in \cite{Ott_Chaos}.

\bibitem{Blakely_QE} J. N. Blakely, L. Illing, and D. J. Gauthier, IEEE J. Quantum Electron. {\bf 40}, 299 (2004). 


\bibitem{Gauthier_PRE} D. J. Gauthier, D. W. Sukow, H. M. Concannon, and J. E. S. Socolar, Phys. Rev. E {\bf 50}, 2343 (1994). 

\bibitem{ETDAS} J. E. S. Socolar, D. W. Sukow, and D. J. Gauthier,  Phys. Rev. E {\bf 50}, 3245 (1994).

\bibitem{Michiels} W. Michiels, K. Engelborghs, V. Vansevenant, and D. Roose,  Automatica {\bf 38}, 747 (2002).

\bibitem{Corron_PRL} K. Myneni, T. A. Barr, N. J. Corron, and S. D. Pethel, Phys. Rev. Lett. {\bf 83}, 2175 (1999).

\bibitem{Wise} F. \"{O}. Ilday, {\it et al.}, Opt. Lett. {\bf 28}, 1365 (2003)


\end{thebibliography}
\end{document}